\input{aipcheck}

\documentclass[
    ,final            
  ]
  {aipproc}

\layoutstyle{6x9}

\begin{document}

\title{Constraint on $\Delta g(x)$ from $\pi^0$ production at RHIC}


\author{M. Hirai}{
address={Department of Physics, Tokyo Institute of Technology \\
2-12-1, Ookayama, Meguro-ku,Tokyo, 152-8550, Japan}}

\author{S. Kumano}{
address={Institute of Particle and Nuclear Studies,
High Energy Accelerator Research Organization (KEK)
1-1, Ooho, Tsukuba, Ibaraki, 305-0801, Japan},
altaddress={Department of Particle and Nuclear Studies,
Graduate University for Advanced Studies 1-1, Ooho, 
Tsukuba, Ibaraki, 305-0801, Japan}}

\author{N. Saito}{
address={Institute of Particle and Nuclear Studies, 
High Energy Accelerator Research Organization (KEK)
1-1, Ooho, Tsukuba, Ibaraki, 305-0801, Japan},
altaddress={Department of Particle and Nuclear Studies,
Graduate University for Advanced Studies 1-1, Ooho, 
Tsukuba, Ibaraki, 305-0801, Japan}}

\begin{abstract}
We determine the polarized gluon distribution $\Delta g(x)$ 
by a global analysis using current DIS and $\pi^0$ asymmetry data.
The $\pi^0$ data from RHIC-Spin experiments provide a strong constraint
on $\Delta g(x)$, so that its uncertainty is reduced.
However, a sign problem appears in the analysis using $\pi^0$ data,
which means that positive and negative distributions are allowed
for $\Delta g(x)$. These two types of solutions are discussed.
\end{abstract}

\maketitle


\section{Introduction}
For investigating the spin structure in the nucleon, 
quark and gluon spin components are determined by global analysis using the asymmetry data
from deeply inelastic scattering (DIS) experiments \cite{polPDFs, aac}. 
The quark component $\Delta \Sigma$ is determined well by present experimental data; 
however, the gluon component $\Delta g$ is not obtained with enough accuracy and still has large uncertainty.
Difficulty of the determination of $\Delta g$ is caused by narrow $Q^2$ rage of polarized DIS data 
compared with the unpolarized data, 
because the polarized gluon distribution $\Delta g(x,Q^2)$
contributes to the structure function $g_1(x,Q^2)$ via $Q^2$ evolution given by the DGLAP equation.
Therefore, we need experimental data sensitive to the gluon distribution,
for example, in hadron, jet, and prompt photon production of polarized 
proton-proton collisions.
Fortunately, the asymmetry data of $\pi^0$ production are measured by the {\sc Phenix} collaboration 
at RHIC \cite{phenix-pi-06}.
Since impact of the data on determination of $\Delta g$ is interesting,
we perform a global analysis with DIS and $\pi^0$ asymmetry data 
and estimate uncertainties of the polarized parton distribution functions (polarized PDFs) \cite{aac06}.

\section{AAC global analysis and results}
For discussing the impact of the $\pi^0$ data, we perform two analyses. 
One is the analysis using only the DIS data, 
and the other uses the DIS and $\pi^0$ asymmetry data.
In these analyses, we assume two constraint conditions. 
One is positivity condition.
It is required as constraint on the large-$x$ behavior of the polarized PDFs 
because the obtained DIS asymmetry tends to exceed one 
due to low accuracy of the experimental data in the large-$x$ region.
The other condition is $SU(3)_f$ flavor symmetry for the polarized antiquark distributions 
because of the low accuracy for flavor separation of the antiquark distributions.
Moreover, the current $\pi^0$ data are insensitive to flavor structure of the polarized PDFs 
because the precise data exist in the low-$p_T$ region 
where the gluon-gluon scattering process dominates.
In our analysis,
the DIS asymmetry are calculated in next-to-leading order (NLO) of $\alpha_s$,
and NLO cross sections of the $\pi^0$ production are estimated by using the K factor method.
Uncertainties of the polarized PDFs are estimated by the Hessian method.
The polarized PDFs are optimized by $\chi^2$ analysis, 
and values of $\chi^2/d.o.f.$ are 0.904 for only DIS data 
and 0.891 for DIS$+\pi^0$ data.

\begin{figure}[b]
  \includegraphics[width=60mm]{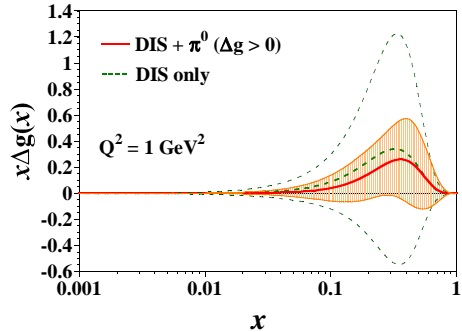}
  \caption{Polarized gluon distributions from the analyses with only DIS and adding $\pi^0$ data.
Solid curve is $x\Delta g(x)$ for DIS$+\pi^0$, and dotted curve is one for only DIS data.
Covered areas by upper and lower curves are these uncertainties.}
  \label{fig:xdg-1&2}
\end{figure}

Figure \ref{fig:xdg-1&2} shows the comparison of the gluon distributions
between two analyses.
While the distribution itself varies slightly, its uncertainty is significantly reduced.
Obtained $\Delta g$ is $0.47\pm1.08$ for the only DIS data, 
and $0.31\pm0.32$ for the DIS$+\pi^0$ data.
It is clear that the uncertainty of $\Delta g$ is significantly reduced 
due to adding the $\pi^0$ data.
This fact implies that the $\pi^0$ data has great impact on the determination of $\Delta g(x)$ 
in comparison with the present DIS data.

\begin{figure}[t]
  \includegraphics[width=60mm]{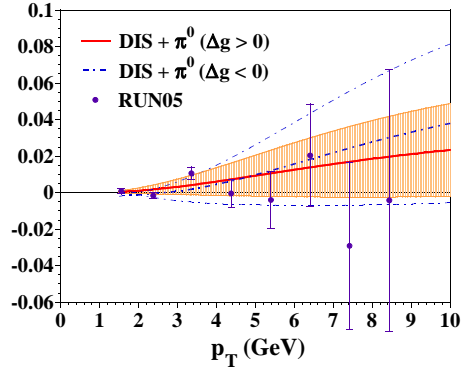}
  \caption{
Comparison of the $\pi^0$ asymmetries between two solutions. 
Experimental data are measured by the {\sc Phenix} collaboration at $\sqrt s=200$ GeV \cite{phenix-pi-06}.}
  \label{fig:asympi0}
\end{figure}

There is, however, a problem in the analysis using the $\pi^0$ data.
Although above results are obtained by assuming the positive distribution for $\Delta g(x)$,
negative distribution can be allowed as a solution of $\Delta g(x)$.
This is because that 
the $gg$ scattering process dominates in the low-$p_T$ region, 
so that the polarized cross section is roughly proportional to square of the gluon distribution:
$\Delta \sigma \propto [\Delta g(x)]^2$.
Therefore, there are two solutions: positive and negative solutions.
In practice, we perform the analysis assuming negative gluon input at initial scale.
The value of minimized $\chi^2$ of the $\pi^0$ data is 11.05 for eight data points, 
and it is almost the same as the positive solution which is 11.18.
Next, the obtained asymmetries are shown in Fig. \ref{fig:asympi0}.
The solid curve is the asymmetry of the positive solution, 
and the dotted-dashed is one of the negative solution.
Although the positive solution passes through the region between 
the second and third data points, 
the negative solution becomes small negative to fit the second data point.
There are not large difference between the $\chi^2$ values 
and the behavior of the asymmetry for both solutions; 
therefore, we cannot choose a solution at this stage.

\begin{figure}[b]
  \includegraphics[width=60mm]{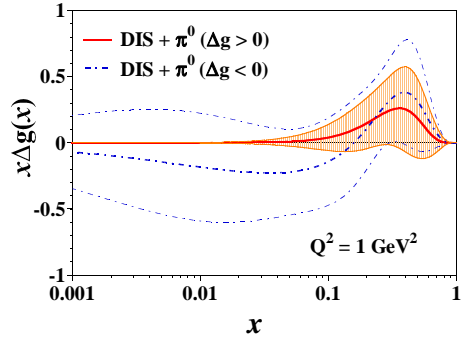}
  \caption{Comparison between the polarized gluon distributions 
           of the positive and negative solutions.}
  \label{fig:xdg-23}
\end{figure}
Negative asymmetry caused by functional forms of the gluon distribution is discussed 
in Ref. \cite{negative-dg}, where a distribution with a node is suggested.
We find indeed such a distribution as shown in Fig. \ref{fig:xdg-23}.
As a negative solution, 
the distribution has a node at $x=0.16$ and becomes negative in the small-$x$ region.
Its uncertainty is large in this small-$x$ region.
The values of $\Delta g$ is $-0.56\pm2.16$. 
It indicates negative value, 
however it has large uncertainty coming from the ambiguity of the small-$x$ behavior 
as shown in Fig. \ref{fig:xdg-23}.

\begin{figure}
  \includegraphics[width=60mm]{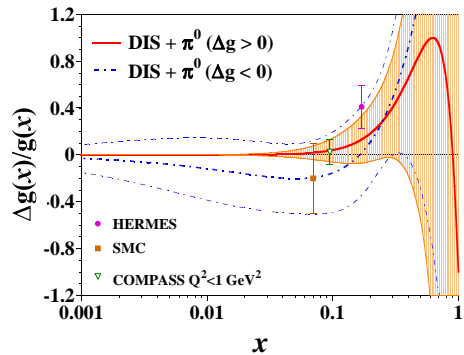}
  \caption{Comparison between the ratios $\Delta g(x)/g(x)$
           for the positive and negative solutions.}
  \label{fig:dgog}
\end{figure}

For discussing reliability of the $\Delta g$ determination, 
we estimate the $\Delta g$ in the following $x$ range; $0.1 \le x \le 1$.
These values are almost the same: $0.30\pm0.32 (\Delta g>0)$ and $0.32\pm0.42 (\Delta g<0)$.
Next, the comparison of the polarization of the gluon distribution is shown in Fig. \ref{fig:dgog}.
Experimental data are measured by di-hadron production in semi-inclusive DIS process.
Although these data are not included in the analysis,
two solutions are consistent with the data within these uncertainties. 
Furthermore these curves seem to be same behavior in the limited $x$ range.
The range is covered by the DIS and $\pi^0$ data; 
therefore, similar solutions are obtained in this rage.
The ambiguity of $\Delta g$ comes from no constraint of the $\pi^0$ data in the small-$x$ region.
Note that analysis using hadron production has ambiguity of fragmentation function sets. 
We should take care of the ambiguity for determination of $\Delta g(x)$.
In this sense, jet production plays an important role in the determination without the ambiguity
because the fragmentation functions are not needed to calculation of jet cross sections.

\section{Summary}
In order to determine the polarized gluon distribution, 
we performed global analysis using present DIS and $\pi^0$ asymmetry data.
Although the uncertainty of the gluon distribution is reduced by adding the $\pi^0$ data,
the sign problem occurs.
The positive and negative $\Delta g$ are allowed at this stage.
However, we obtain consistent results for two solutions in the region $x>0.1$ 
which is covered by the present experimental data.
The difference of the sign comes from the extrapolated behavior of
the gluon distribution in the smaller-$x$ region. 
In the region, there is no constraint on the gluon distribution.
In order to reduce the large uncertainty, 
experimental data covering a wide rage of $x$ are required.
In particular, the data in low $p_T$ at $\sqrt s=500$ GeV will provide
a constraint on the small-$x$ behavior.

\begin{theacknowledgments}
The authors were supported by 
the Grant-in-Aid for Scientific Research from the Japan Ministry of Education, Culture, Sports,
Science, and Technology.
\end{theacknowledgments}



\bibliographystyle{aipproc}   


\begin{thebibliography}{9}
\newcommand{\etal}{{\it et al.}}
\bibitem{polPDFs} 
M. Gl\"uck, E. Reya, M. Stratmann, and W. Vogelsang,
                       Phys. Rev. {\bf D63}, 094005 (2001);
J. Bl\"umlein and H. B\"ottcher,
                       Nucl. Phys. {\bf B636}, 225 (2002);
D. de Florian, G. A. Navarro, and R. Sassot,
                       Phys. Rev. {\bf D71},  094018 (2005);
E. Leader, A. V. Sidorov, and D. B. Stamenov,
                       Phys. Rev. {\bf D73}, 034023 (2006).

\bibitem{aac} Asymmetry Analysis Collaboration (AAC),
Y.~Goto, \etal,Phys. Rev. {\bf D62}, 034017 (2000);
M.~Hirai, S.~Kumano, and N.~Saito, Phys. Rev. {\bf D69}, 054021 (2004).

\bibitem{phenix-pi-06} {\sc Phenix} collaboration, 
                       K. Boyle,  talk at the XVIIth Particles
                       and Nuclei International Conference (PANIC05),
                       http://www.panic05.lanl.gov/ .

\bibitem{aac06} AAC, M.~Hirai, S.~Kumano, N.~Saito , 
Phys. Rev. {\bf D74}, 014015 (2006).

\bibitem{negative-dg} 
B.~J\"ager, M.~Stratmann, S.~Kretzer, and W.~Vogelsang,
                     Phys. Rev. Lett. {\bf 92}, 121803 (2004);
M.~Hirai and K.~Shdoh, Phys. Rev. {\bf D71},  014022 (2005).


\end{thebibliography}




\end{document}